\documentclass[11pt]{article}
\usepackage{graphicx}
\usepackage{amssymb}
\usepackage{amsmath}
\usepackage{graphicx}
\usepackage{amstext}
\usepackage{esint}
\usepackage[utf8]{inputenc}
\usepackage{color}
\usepackage{epsfig}
\usepackage{epstopdf}

\renewcommand{\vec}[1]{{\bf #1}}       %%%  vectors in bold
\def\beq{\begin{eqnarray}}    %%%  begequation/eqnarray
\def\eeq{\end{eqnarray}}      %%%  endequation/eqnarray
\newcommand{\rL}{\rho_\Lambda}

\newcommand{\CC}{\Lambda}

\newcommand{\rmr}{\rho_m}

\newcommand{\rLo}{\rho_{\CC 0}}

\begin{document}
%\pubblock

%\today

%\vspace{1cm}

 \hyphenation{nu-cleo-syn-the-sis u-sing si-mu-la-te ma-king
cos-mo-lo-gy know-led-ge e-vi-den-ce stu-dies be-ha-vi-or
res-pec-ti-ve-ly appro-xi-ma-te-ly gra-vi-ty sca-ling
ge-ne-ra-li-zed con-si-de-red}

%%%%%%%%%%%%%%%%%%%%%%%%%%%%%%%%%%%%%%%%%%%%%%%%%%%%%%%%%

%\newpage

%%%%%%%%%%%%%%%%%%%%%%%%%%%%%%%%%%%%%%%%%%%%%%%%%%%%%%%%%
%\flushright{UB-ECM-PF-06/24 }\\

\begin{center}
\textit{\LARGE Dynamical dark energy: scalar fields and running vacuum} \vskip 2mm

 \vskip 8mm

\textbf{Joan Sol\`a, Adri\`a G\'omez-Valent, and Javier de Cruz P\'erez}

\vskip 0.5cm
Departament de F\'isica Qu\`antica i Astrof\'isica, and Institute of Cosmos Sciences,\\ Universitat de Barcelona, \\
Av. Diagonal 647, E-08028 Barcelona, Catalonia, Spain

\vskip0.5cm

\vskip0.4cm

E-mails:   sola@fqa.ub.edu, adriagova@fqa.ub.edu, decruz@fqa.ub.edu

 \vskip2mm

\end{center}
\vskip 15mm

\begin{quotation}
\noindent {\large\it \underline{Abstract}}.
Recent analyses in the literature suggest that the concordance $\CC$CDM model with rigid cosmological term, $\CC=$const., may not be the best description of the cosmic acceleration. The class of ``running vacuum models'', in which $\CC=\CC(H)$ evolves with the Hubble rate, has been shown to fit the string of SNIa+BAO+$H(z)$+LSS+CMB data significantly better than the $\CC$CDM. Here we provide further evidence on the time-evolving nature of the dark energy (DE) by fitting the same cosmological data in terms of scalar fields. As a representative model we use the original Peebles \& Ratra potential, $V\propto\phi^{-\alpha}$. We find clear signs of dynamical DE at  $\sim 4\sigma$ c.l., thus reconfirming through a nontrivial scalar field approach the strong hints formerly found with other models and parametrizations.
\end{quotation}
\vskip 5mm

\newpage

%\tableofcontents

\newpage

%%%%%%%%%%%%%%%%%%%%%%%%%%%%%%%%%%%%%%%%%%%%%%%%%%%%%%%%%%%%%%%%%
%%%%%%%%%%%%%%%%%%%%%%%%%%%%%%%%%%%%%%%%%%%%%%%%%%%%%%%%%%%%%%%%%
%%%%%%%%%%%%%%%%%%%%%%%%%%%%%%%%%%%%%%%%%%%%%%%%%%%%%%%%%%%%%%%%%

\section{Introduction}\label{intro}

It is an observational fact that the Universe is in accelerated expansion\,\cite{SNIaRiess,SNIaPerl}. The agent responsible for it has not been precisely identified, it is generically called the dark energy (DE). The simplest hypothesis for the DE is to associate it with a positive cosmological constant (CC) in Einstein's equations, denoted $\CC$. This is at least the picture advocated by the standard or ``concordance'' $\CC$CDM model, which is fairly consistent with numerous observations\,\cite{Planck2015}.

However, the $\CC$-term is usually associated with the vacuum energy density $\rL=\CC/(8\pi G)$ ($G$ being Newton's gravitational coupling) and this leads to the so-called cosmological constant (CC) problem\,\cite{Weinberg,Padmanabhan2003,PeeblesRatra2003,CopeSamiTsuji2006}, which appears because the prediction for  $\CC$ in quantum field theory (QFT) differs from the measured value of $\rL$ by many orders of magnitude\,\cite{SNIaRiess,SNIaPerl,Planck2015}. For this reason it has been suggested that it would help to alleviate such problem, including the associated cosmic coincidence problem\,\cite{PeeblesRatra88,RatraPeebles88,Steindardt97}, if the DE would be dynamical, i.e. slowly evolving with the cosmic expansion. This could be achieved e.g. through scalar field models. They have been proposed in the past either to adjust dynamically the value of the vacuum energy\,\cite{Dolgov83} in different versions (e.g. the cosmon\,\cite{PSW}) or to endow the DE of a convenient dynamics with the notion of quintessence, etc. \cite{PeeblesRatra88,RatraPeebles88,Wetterich88,Wetterich95,Caldwell98,Amendola2000}, see also\,\cite{Padmanabhan2003,PeeblesRatra2003,CopeSamiTsuji2006,DEBook}. Many other proposals have been considered, in particular the possibility that the $\CC$-term acquires some phenomenological ad hoc evolution with the cosmic time\,\cite{CCtime}.
Ultimately this matter should be settled empirically, and, interestingly enough, this is possible at present owing to the rich wealth of cosmological data at our disposal.

For example, the possibility of a dynamical vacuum energy density slowly evolving with the Hubble rate, $\rL=\rL(H)$, has recently been explored in detail. Some of these models, particularly the class of running vacuum models (RVM's)  (cf. e.g. the reviews \cite{JSPRev2013,SolGom2015,JSPRev2016}), have been put to the test. These models have an interesting theoretical motivation grounded in QFT in curved spacetime. Although a general Lagrangian formulation comparable to that of scalar fields has not been found at the moment, a non-local effective action description can be obtained in certain cases \,\cite{JSPRev2013,Fossil07}. Phenomenologically, the RVM's have recently been carefully confronted againts observations and with significant success. For example, the analysis of \cite{ApJL} reveals that they fit better the cosmological data than the $\CC$CDM at a confidence level of around $3\sigma$. This significance has recently been promoted to $4\sigma$ in \cite{RVM,DVM}, see also \cite{JSPRev2016} for a summary. The next natural question that can be formulated is whether the traditional class of $\phi$CDM models, which do have a local Lagrangian description, and in which the DE is described in terms of a scalar field $\phi$ with some standard form for its potential $V(\phi)$, are also capable of capturing clear signs of dynamical DE using the same set of cosmological observations used for fitting the RVM.

We devote this Letter to show that, indeed, it is so. We compare these two kind of different models and also with the results obtained using the well-known XCDM\,\cite{XCDM} and CPL\cite{CPL1,CPL2,CPL3} parametrizations of the DE. The upshot is that we are able to collect further evidence on the time evolution of the DE from different types of models at a confidence level of $\sim 4\sigma$. This result is very encouraging and suggests that the imprint of dynamical DE in the modern data is fairly robust and can be clearly decoded using independent formulations.

%%%%%%%%%%%%%%%%%%%%%%%%%%%%%%%%%%%%%%%%%%%%%%%%%%%%%%%%%%%%%%%%%
%%%%%%%%%%%%%%%%%%%%%%%%%%%%%%%%%%%%%%%%%%%%%%%%%%%%%%%%%%%%%%%%%
%%%%%%%%%%%%%%%%%%%%%%%%%%%%%%%%%%%%%%%%%%%%%%%%%%%%%%%%%%%%%%%%%

\newpage
\section{$\phi$CDM with Peebles \& Ratra potential}\label{sect:phiCDM}
Suppose that the dark energy is described in terms of some scalar field $\phi$ with a standard form for its potential $V(\phi)$, see below. We wish to compare its ability to describe the data with that of the $\CC$CDM, and also with other models of DE existing in the literature. The data used in our analysis will be the same one used in our previous studies\,\cite{ApJL,RVM,DVM} namely data on the distant supernovae (SNIa), the baryonic acoustic oscillations (BAO), the Hubble parameter at different redshifts, $H(z_i)$, the large scale structure formation data (LSS), and the cosmic microwave background (CMB). We denote this string of data as SNIa+BAO+$H(z)$+LSS+CMB.  Precise information on these data and corresponding observational references are given in the aforementioned papers and are also summarized in the caption of Table 1. The main results of our analysis are displayed in Tables 1-3 and Figures 1-5, which we will account in detail throughout our exposition.

We start by explaining our theoretical treatment of the $\phi$CDM model in order to optimally confront it with observations. The scalar field $\phi$ is taken to be dimensionless, being its energy density and pressure given by
\begin{equation}\label{eq:rhophi}
\rho_\phi=\frac{M^2_{pl}}{16\pi}\left[\frac{\dot{\phi}^2}{2}+V(\phi)\right]\,,\ p_\phi=\frac{M^2_{pl}}{16\pi}\left[\frac{\dot{\phi}^2}{2}-V(\phi)\right]\,.
\end{equation}
Here $M_{pl}=1/\sqrt{G}=1.22\times 10^{19}$ GeV is the Planck mass, in natural units.
As a representative potential we adopt the original Peebles \& Ratra (PR) form\,\cite{PeeblesRatra88}:
\begin{equation}\label{eq:PRpotential}
V(\phi)=\frac{1}{2}\kappa M_{pl}^2\phi^{-\alpha}\,,
\end{equation}
in which $\kappa$ and $\alpha$ are dimensionless parameters. These are to be determined in our fit to the overall cosmological data. The motivation for such potential is well described in the original paper \cite{PeeblesRatra88}. In a nutshell: such potential stands for the power-law tail of a more complete effective potential in which inflation is also comprised. We expect $\alpha$ to be positive and sufficiently small such that $V(\phi)$ can mimic an approximate CC that is decreasing slowly with time, in fact more slowly than the matter density. Furthermore, we must have $0<\kappa\ll 1$  such that $V(\phi)$ can be positive and of the order of the measured value $\rLo\sim 10^{-47}$ GeV$^4$. In the late Universe the tail of the mildly declining potential finally surfaces over the matter density (not far away in our past) and appears as an approximate CC which dominates since then.  Recent studies have considered the PR-potential in the light of the cosmological data, see e.g. \cite{Pavlov2014,Avsajanishvili2014,Farooq2013,Pourtsidou1,Pourtsidou2}. Here we show that the asset of current observations indicates strong signs of dynamical DE which can be parametrized with such potential. In this way, we corroborate the unambiguous signs recently obtained with independent DE models\,\cite{ApJL,RVM,DVM} and with a similar level of confidence.

%%%%%%%%%%%%%%%%%%%%%%%%%%%%%%%%%%%%%%%%%%%%%%%%%%%%%%%%%%%%%%%%%%%%%%%%%%%%%%%%%%%%%%%%%%%%%
\begin{table*}
\begin{center}
\begin{scriptsize}
\resizebox{1\textwidth}{!}{
\begin{tabular}{| c | c |c | c | c | c |}
\multicolumn{1}{c}{Model} &  \multicolumn{1}{c}{$\Omega_m$} &  \multicolumn{1}{c}{$\omega_b= \Omega_b h^2$} & \multicolumn{1}{c}{{\small$n_s$}}  &  \multicolumn{1}{c}{$h$} &
\multicolumn{1}{c}{$\chi^2_{\rm min}/dof$}
\\\hline
{$\Lambda$CDM} & $0.294\pm 0.004$ & $0.02255\pm 0.00013$ &$0.976\pm 0.003$& $0.693\pm 0.004$ & 90.44/85    \\
\hline
 \end{tabular}
 }
\caption{{\scriptsize The best-fit values for the $\CC$CDM parameters $(\Omega_m,\omega_b,n_s,h)$. We use a total of $89$ data points from SNIa+BAO+$H(z)$+LSS+CMB observables in our fit: namely $31$ points from the JLA sample of SNIa\,\cite{BetouleJLA}, $11$ from BAO\,\cite{Beutler2011,Ross2015,Kazin2014,GilMarin2016,Delubac2015,Aubourg2015}, $30$  from $H(z)$\,\cite{Zhang2014,Jimenez2003,Simon2005,Moresco2012,Moresco2016,Stern2010,Moresco2015}, $13$ from linear growth \cite{GilMarin2016,Beutler2012,Feix2015,Simpson2016,Blake2013,Blake2011BAO,Springob2016,Granett2015,Guzzo2008,SongandPercival2009}, and $4$ from CMB\,\cite{Huang2015}. For a summarized description of these data, see \cite{RVM}. The quoted number of degrees of freedom ($dof$) is equal to the number of data points minus the number of independent fitting parameters ($4$ for the $\CC$CDM). For the CMB data we have used the marginalized mean values and standard deviation for the parameters of the compressed likelihood for Planck 2015 TT,TE,EE + lowP data from \cite{Huang2015}. The parameter M in the SNIa sector\,\cite{BetouleJLA} was dealt with as a nuisance parameter and has been marginalized over analytically. The best-fit values and the associated uncertainties for each parameter in the table have been obtained by numerically marginalizing over the remaining parameters\,\cite{AmendolaStatistics}.}}
 \end{scriptsize}
\end{center}
\label{tableFit1}
\end{table*}
The scalar field of the $\phi$CDM models satisfies the Klein-Gordon equation in the context of the Friedmann-Lema\^\i tre-Robertson-Walker (FLRW) metric: $\ddot{\phi}+3H\dot{\phi}+{dV}/{d\phi}=0$, where $H=\dot{a}/a$ is the Hubble function.
In some cases the corresponding solutions possess the property of having an attractor-like behavior, in which a large family of solutions are drawn towards a common trajectory\,\cite{RatraPeebles88,SteinWangIvaylo09a,SteinWangIvaylo09b}. If there is a long period of convergence of all the family members to that common trajectory, the latter is called a ``tracker solution''\,\cite{SteinWangIvaylo09a,SteinWangIvaylo09b}. When the tracking mechanism is at work, it funnels a large range of initial conditions into a common final state for a long time (or forever, if the convergence is strict). Not all potentials $V$ admit tracking solutions, only those fulfilling the ``tracker condition'' $\Gamma\equiv V\,V''/(V')^2>1$\,\cite{SteinWangIvaylo09a,SteinWangIvaylo09b}, where $V'=\partial V/\partial\phi$.  For the Peebles-Ratra potential (\ref{eq:PRpotential}), one easily finds $\Gamma=1+1/\alpha$, so it satisfies such condition precisely for $\alpha>0$.

{It is frequently possible to seek power-law solutions, i.e. $\phi(t)=A\,t^p$,
for the periods when the energy density of the Universe is dominated by some conserved matter component} $\rho(a)=\rho_1\left({a_1}/{a}\right)^n$ (we may call these periods ``nth-epochs''). For instance,
$n=3$ for the matter-dominated epoch (MDE) and $n=4$ for the radiation-dominated epoch (RDE), with
$a_1$ the scale factor at some cosmic time $t_1$ when the corresponding component dominates. We define  $a=1$ as the current value. Solving Friedmann's equation in flat space,
$3H^2(a)=8\pi G\rho(a)$, we find  $H(t)={2}/(nt)$ as a function of the cosmic time in the nth-epoch. Substituting these relations in the  Klein-Gordon equation with the Peebles-Ratra potential (\ref{eq:PRpotential}) leads to
\begin{equation}
p=\frac{2}{\alpha+2}\,,\qquad A^{\alpha+2}=\frac{\alpha(\alpha+2)^2M_{pl}^2\kappa n}{4(6\alpha+12-n\alpha)}\,.
\end{equation}
From the power-law form we find the evolution of the scalar field with the cosmic time:
\begin{equation}\label{eq:initialphi}
\phi(t)=\left[\frac{\alpha(\alpha+2)^2M_{pl}^2\kappa n}{4(6\alpha+12-n\alpha)}\right]^{1/(\alpha+2)}t^{2/(\alpha+2)}\,.
\end{equation}
In any of the nth-epochs the equation of state (EoS) of the scalar field remains stationary. A straightforward calculation from (\ref{eq:rhophi}), (\ref{eq:PRpotential}) and (\ref{eq:initialphi}) leads to a very compact form for the EoS:
\begin{equation}\label{eq:EoSphi}
w_\phi=\frac{p_{\phi}}{\rho_{\phi}}=-1+\frac{\alpha n}{3(2+\alpha)}\,.
\end{equation}
Since the matter EoS in the nth-epoch is given by $\omega_n=-1+n/3$, it is clear that (\ref{eq:EoSphi}) can be rewritten also as $w_\phi=(\alpha\omega_n-2)/(\alpha+2)$. This is precisely the form predicted by the tracker solutions\,\cite{SteinWangIvaylo09a,SteinWangIvaylo09b}, in which the condition $w_\phi<\omega_n$ is also secured since $|\alpha|$ is expected small. In addition, $w_\phi$ remains constant in the RDE and MDE, but its value does \textit{not} depend on $\kappa$, only on $n$ (or $\omega_n$) and $\alpha$. The fitting analysis presented in Table 2 shows that $\alpha={\cal O}(0.1)>0$ and therefore $w_\phi\gtrsim-1$. It means that the scalar field behaves as quintessence in the pure RD and MD epochs (cf. the plateaus at constant values  $w_\phi\gtrsim-1$ in Fig.\,1). Notice that the behavior of $w_\phi$ in the interpolating epochs, including the period near our time, is \textit{not} constant (in contrast to the XCDM, see next section) and requires numerical solution of the field equations. See also\, \cite{PodariuRatra} for related studies.

We can trade the cosmic time in (\ref{eq:initialphi}) for the scale factor. This is possible using $t^2=3/(2\pi G n^2\rho)$ (which follows from Friedmann's equation in the nth-epoch) and  $\rho(a)=\rho_{0}a^{-n}=\rho_{c0}\,\Omega\,a^{-n}$, where $\Omega=\Omega_m, \Omega_r$ are the present values of the cosmological density parameters for matter ($n=3$) or radiation ($n=4$) respectively, with $\rho_{c 0}=3H_0^2/(8\pi\,G)$  the current critical energy density. Notice that $\Omega_m=\Omega_{dm}+\Omega_b$ involves both dark matter and baryons. In this way we can determine $\phi$ as a function of the scale factor in the nth-epoch. For example, in the MDE we obtain

\begin{equation}\label{eq:Phia}
\phi(a)=\left[\frac{\alpha(\alpha+2)^2\bar{\kappa}}{9\times 10^4\omega_m(\alpha+4)}\right]^{1/(\alpha+2)}a^{3/(\alpha+2)}\,.
\end{equation}

Here we have conventionally defined the reduced matter density parameter $\omega_m\equiv\Omega_m\,h^2$, in which the reduced Hubble constant $h$ is defined as usual from $H_0\equiv 100h\,\varsigma$, with $\varsigma\equiv 1 Km/s/Mpc=2.133\times10^{-44} GeV$ (in natural units). Finally, for convenience we have introduced in (\ref{eq:Phia}) the dimensionless parameter $\bar{\kappa}$ through $\kappa\,M_P^2\equiv \bar{\kappa}\,\varsigma^2$.

%%%%%%%%%%%%%%%%%%%%%%%%%%%%%%%%%%%%%%%%%%%%%%%%%%%%%%%%%%%%%%%%%%%%%%%%%%%%%%%%%%%%%%%%%%%%

\begin{table*}
\begin{center}
\begin{scriptsize}
\resizebox{1\textwidth}{!}{
\begin{tabular}{| c | c |c | c | c | c | c | c | c | c|c|}
\multicolumn{1}{c}{Model} &  \multicolumn{1}{c}{$\omega_m=\Omega_m h^2$} &  \multicolumn{1}{c}{$\omega_b=\Omega_b h^2$} & \multicolumn{1}{c}{{\small$n_s$}}  &  \multicolumn{1}{c}{$\alpha$} &  \multicolumn{1}{c}{$\bar{\kappa}$}&  \multicolumn{1}{c}{$\chi^2_{\rm min}/dof$} & \multicolumn{1}{c}{$\Delta{\rm AIC}$} & \multicolumn{1}{c}{$\Delta{\rm BIC}$}\vspace{0.5mm}
\\\hline
$\phi$CDM  &  $0.1403\pm 0.0008$& $0.02264\pm 0.00014 $&$0.977\pm 0.004$& $0.219\pm 0.057$  & $(32.5\pm1.1)\times 10^{3}$ &  74.85/84 & 13.34 & 11.10 \\
\hline
 \end{tabular}
 }
\caption{{\scriptsize The best-fit values for the parameter fitting vector (\ref{eq:vfittingPhiCDM}) of the $\phi$CDM model with Peebles \& Ratra potential (\ref{eq:PRpotential}), including their statistical significance ($\chi^2$-test and Akaike and Bayesian information criteria, AIC and BIC, see the text). We use the same cosmological data set as in Table 1. The large and positive values of $\Delta$AIC and $\Delta$BIC strongly favor the $\phi$CDM model against the $\CC$CDM.  The specific $\phi$CDM fitting parameters are $\bar{\kappa}$ and $\alpha$. The remaining parameters  $(\omega_m,\omega_b,n_s)$ are standard (see text).  The number of independent fitting parameters is $5$, see Eq.\,(\ref{eq:vfittingPhiCDM})-- one more than in the $\CC$CDM. Using the best-fit values and the overall covariance matrix derived from our fit, we obtain: $h=0.671\pm 0.006$ and $\Omega_m=0.311\pm 0.006$, which allows direct comparison with Table 1. We find $\sim 4\sigma$ evidence in favor of $\alpha>0$. Correspondingly the EoS of $\phi$ at present appears quintessence-like at $4\sigma$ confidence level: $w_\phi= -0.931\pm 0.017$.}}
 \end{scriptsize}
\end{center}
\label{tableFit2}
\end{table*}

Equation (\ref{eq:Phia}) is convenient since it is expressed in terms of the independent parameters that enter our fit, see below.  Let us note  that $\phi(a)$, together with its derivative $\phi^{\prime}(a)=d\phi(a)/da$, allow us to fix the initial conditions in the MDE (a similar expression can be obtained for the RDE). Once these conditions are settled analytically we have to solve numerically the Klein-Gordon equation, coupled to the cosmological equations, to obtain the exact solution. Such solution must, of course, be in accordance with (\ref{eq:Phia}) in the pure MDE. The exact EoS is also a function $w_\phi=w_\phi(a)$, which coincides with the constant value (\ref{eq:EoSphi}) in the corresponding nth-epoch, but interpolates nontrivially between them. At the same time it also interpolates between the MDE and the DE-dominated epoch in our recent past, in which the scalar field energy density surfaces above the nonrelativistic matter density, i.e. $\rho_{\phi}(a)\gtrsim \rho_m(a)$, at a value of $a$ near the current one $a=1$. %(which we normalize to $1$).
The plots for the deceleration parameter, $q=-\ddot{a}/aH^2$, and the scalar field EoS, $w_\phi(a)$, for the best fit parameters of Table 2 are shown in Fig. 1.  The transition point from deceleration to acceleration ($q=0$) is at $z_t=0.628$, {which is in good agreement with the values obtained in other works \cite{FarooqRatra,Farooqztr2016}}, and is also reasonably near the $\CC$CDM one ($z_t^{\CC{\rm CDM}}=0.687$) for the best fit values in Tables 1 and 2. The plots for $\phi(a)$ and the energy densities are displayed in Fig. 2. 
From equations  (2) and (6) we can see that in the early MDE the potential of the scalar field decays as $V\sim a^{-3\alpha/(2+\alpha)}\sim  a^{-3\alpha/2} $, where in the last step we used the fact that  $\alpha$ is small. Clearly the decaying behavior of $V$ with the expansion is much softer than that of the matter density,  $\rho_m\sim a^{-3}$, and for this reason the DE density associated to the scalar field does not play any role until we approach the current time. This fact is apparent in Fig. 2 (right), where we numerically plot the dimensionless density parameters $\Omega_i(a)=\rho_i(a)/\rho_c(a)$ as a function of the scale factor, where $\rho_c(a)=3H^2(a)/(8\pi G)$ is the evolving critical density.

{As indicated above, the current value of the EoS} can only be known after numerically solving the equations for the best fit parameters in Table 2, with the result $w_\phi(z=0)= -0.931\pm 0.017$ (cf. Fig. 1). Such result lies clearly in the quintessence regime and with a significance of $4\sigma$. It is essentially consistent with the dynamical character of the DE derived from the non-vanishing value of $\alpha$ in Table 2.

In regard to the value of $h$, there is a significant tension between non-local measurements of $h$, e.g. \cite{Planck2015,Aubourg2015,ChenRatra,ACT,HuillierShafieloo,BernalVerdeRiess,LukovicAgostinoVittorio}, and local ones, e.g. \cite{RiessH0}. Some of these values can differ by $3\sigma$ or more. For the $\CC$CDM model we find $h=0.693\pm 0.004$ (cf. Table 1), which is in between the ones of \cite{Planck2015} and \cite{RiessH0} and is compatible with the value presented in \cite{WMAP9}. For the $\phi$CDM, our best-fit value is $h=0.671\pm 0.006$ (cf. caption of Table 2), which differs by more than $3\sigma$ with respect to the $\CC$CDM one in our Table 1. Still, both remain perfectly consistent with the recent estimates of $h$ from Hubble parameter measurements at intermediate redshifts\,\cite{Chen2016}. At the moment it is not possible to distinguish models on the sole basis of $H(z)$ measurements. Fortunately, the combined use of the different sorts of SNIa+BAO+$H(z)$+LSS+CMB data offers nowadays a real possibility to elucidate which models are phenomenologically preferred.

%%%%%%%%%%%%%%%%%%%%%%%%%%%%%%%%%%%%%%%%%%%%%%%%%%%%%%%%%%%%%%%%%%%%%%%%%%%%%%%%%%%%%%%%%%%%%%%%

\begin{table*}
\begin{center}
\begin{scriptsize}
\resizebox{1\textwidth}{!}{
\begin{tabular}{| c | c |c | c | c | c | c | c | c | c|c|}
\multicolumn{1}{c}{Model} &  \multicolumn{1}{c}{$\Omega_m$} &  \multicolumn{1}{c}{$\omega_b= \Omega_b h^2$} & \multicolumn{1}{c}{{\small$n_s$}}  &  \multicolumn{1}{c}{$h$} &  \multicolumn{1}{c}{$\nu$}&  \multicolumn{1}{c}{$w_0$} &  \multicolumn{1}{c}{$w_1$} &
\multicolumn{1}{c}{$\chi^2_{\rm min}/dof$} & \multicolumn{1}{c}{$\Delta{\rm AIC}$} & \multicolumn{1}{c}{$\Delta{\rm BIC}$}\vspace{0.5mm}
\\\hline
{\small XCDM} & $0.312\pm 0.007$ & $0.02264\pm 0.00014$ &$0.977\pm 0.004$& $0.670\pm 0.007$ & - & $-0.916\pm 0.021$ & - & 74.91/84 & 13.28 & 11.04\\
\hline
{\small CPL} & $0.311\pm 0.009$ & $0.02265\pm 0.00014$ &$0.977\pm 0.004$& $0.672\pm 0.009$ & - & $-0.937\pm 0.085$ & $0.064\pm 0.247$  & 74.85/83 & 11.04 & 6.61\\
\hline
{\small RVM} & $0.303\pm 0.005$ & $0.02231\pm 0.00015$ &$0.965\pm 0.004$& $0.676\pm 0.005$ & $0.00165\pm 0.00038$ & -1 & - & 70.32/84 & 17.87 & 15.63\\
\hline
 \end{tabular}
 }
\caption{{\scriptsize The best-fit values for the running vacuum model (RVM), together with the XCDM and CPL parametrizations, including also their statistical significance ($\chi^2$-test and Akaike and Bayesian information criteria, AIC and BIC) as compared to the $\CC$CDM (cf. Table 1). We use the same string of cosmological SNIa+BAO+$H(z)$+LSS+CMB data as in Tables 1 and 2. The specific fitting parameters for these models are $\nu,w_0,$ and $(w_0,w_1)$ for RVM, XCDM and CPL, respectively.  The remaining parameters  are standard. For the models RVM and XCDM the number of independent fitting parameters is $5$, exactly as in the $\phi$CDM. For the CPL parametrization there is one additional parameter ($w_1$). The large and positive values of $\Delta$AIC and $\Delta$BIC strongly favor the RVM and XCDM against the $\CC$CDM. The CPL is only moderately favored as compared to the $\CC$CDM and much less favored than the $\phi$CDM, RVM and XCDM.}}
 \end{scriptsize}
\end{center}
\label{tableFit3}
\end{table*}
%%%%%%%%%%%%%%%%%%%%%%%%%%%%%%%%%%%%%%%%%%%%%%%%%%%%%%%%%%%%%%%%%%%%%%%%%%%%%%%%%%%%%%%%%%%%%%%%%%

Let us now describe the computational procedure that we have followed for the $\phi$CDM model. The initial conditions must be expressed in terms of the parameters that enter our fit. These are defined by means of the following $5$-dimensional fitting vector:
\begin{equation}\label{eq:vfittingPhiCDM}
\vec{p}_{\phi{\rm CDM}}=(\omega_m,\omega_b,n_s,\alpha,\bar{\kappa})
\end{equation}
where $\omega_b\equiv\Omega_b\,h^2$ is the baryonic component and $n_s$ is the spectral index. These two parameters are specifically involved in the fitting of the CMB and LSS data ($\omega_b$ enters the fitting of the BAO data too), whereas the other three also enter the background analysis, see \cite{ApJL,RVM,DVM} and \cite{GomSolBas2015,AGVsola2015,AGVsolKa2015} for more details in the methodology. For the $\phi$CDM we have just one more fitting parameter than in the $\CC$CDM, i.e. $5$ instead of $4$ parameters (cf. Tables 1 and 2).
However, in contrast to the $\CC$CDM, for the $\phi$CDM we are fitting the combined parameter $\omega_m=\Omega_m h^2$ rather than $\Omega_m$ and $h$ separately.
The reason is that $h$ (and hence $H_0$) is not a direct fitting parameter in this case since the Hubble function values are determined from Friedmann's equation $3H^2=8\pi\,G(\rho_{\phi}+\rho_m)$, where $\rho_{\phi}$ is given in Eq.\,(\ref{eq:rhophi}) and $\rho_m=\rho_{c 0}\Omega_m a^{-3}=(3\times 10^4/8\pi G)\varsigma^2\,\omega_m\,a^{-3}$ is the conserved matter component. This is tantamount to saying that $h$ is eventually determined from the parameters of the potential and the reduced matter density $\omega_m$. For instance, in the MDE it is not difficult to show that
\begin{equation}\label{eq:barH2}
\bar{H}^2(a)=\frac{\bar{\kappa}\,\phi^{-\alpha}(a)+1.2\times 10^5\,\omega_m\,a^{-3}}{12-a^2\phi^{\prime 2}(a)}\,,
\end{equation}
where we have defined the dimensionless $\bar{H}=H/\varsigma$, and used $\dot{\phi}=a\,H\,\phi^{\prime}(a)$. As we can see from (\ref{eq:barH2}), the value of $h\equiv\bar{H}(a=1)/100$ is determined once the three parameters $(\omega_m,\alpha,\bar{\kappa})$ of the fitting vector (\ref{eq:vfittingPhiCDM}) are given, and then $\Omega_m=\omega_m/h^2$ becomes also determined.  Recall that $\phi(a)$ is obtained by solving numerically the Klein-Gordon equation under appropriate initial conditions (see below) which also depend on the above fitting parameters. As a differential equation in the scale factor, the Klein-Gordon equation reads
\begin{equation}\label{eq:KGa}
\phi^{\prime\prime}+\phi^\prime\left(\frac{\bar{H}^\prime}{\bar{H}}+\frac{4}{a}\right)-\frac{\alpha}{2}\frac{\bar{\kappa}\phi^{-(\alpha+1)}}{(a\bar{H})^2}=0\,.
\end{equation}
It can be solved after inserting (\ref{eq:barH2}) in it, together with
\begin{equation}
\bar{H}^\prime=-\frac{3}{2a\bar{H}}\left(\frac{a^2\bar{H}^2\phi^{\prime 2}}{6}+10^4\,\omega_m a^{-3}\right)\,.
\end{equation}
The last formula is just a convenient rephrasing of the expression $\dot{H}=-4\pi\,G\left(\rho_m+p_m+\rho_{\phi}+p_{\phi}\right)$ upon writing it in the above set of variables. According to (\ref{eq:rhophi}), the sum of density and pressure for $\phi$ reads $\rho_{\phi}+p_{\phi}=\dot{\phi}^2/(16\pi{G})=a^2 \bar{H}^2{\phi'}^2 \varsigma^2/(16\pi{G})$, and  of course $p_m=0$ for the matter pressure after the RDE.

The initial conditions for solving (\ref{eq:KGa}) are fixed in the mentioned nth-epochs of the cosmic evolution. They are determined from the values of the fitting parameters in (\ref{eq:vfittingPhiCDM}). For example if we set these conditions in the MDE they are defined from the expression of $\phi(a)$ in Eq.\,(\ref{eq:Phia}), and its derivative $\phi^{\prime}(a)$, both taken at some point deep in the MDE, say at a redshift $z>100$, i.e. $a<1/100$.  The result does not depend on the particular choice in this redshift range provided we do not approach too much the decoupling epoch  ($z\simeq 1100$), where the radiation component starts to be appreciable. We have also iterated our calculation when we take the initial conditions deep in the RDE ($n=4$), in which the radiation component $\rho_r$ dominates. In this case $\omega_m=\Omega_m h^2$ is replaced by $\omega_r=\Omega_r h^2$, which is a function of the radiation temperature and the effective number of neutrino species, $N_{eff}$. We find the same results as with the initial conditions settled in the MDE. In both cases the fitting values do agree and are those indicated in Table 2. Let us also mention that when we start from the RDE we find that $\rho_{\phi}(a)\ll\rho_r(a)$ at (and around) the time of BBN (Big Bang Nucleosynthesis), where $a\sim 10^{-9}$, thus insuring that the primordial synthesis of the light elements remains unscathed.

Consistency with BBN is indeed a very important point that motivates the Peebles \& Ratra's inverse power potential $\phi$CDM, Eq.\,(\ref{eq:PRpotential}), together with the existence of the attractor solution. Compared, say to the exponential potential, $V(\phi)=V_0\,e^{-\lambda\,\phi/M_P}$, the latter is inconsistent with BBN (if $\lambda$ is too small) or cannot be important enough to cause accelerated expansion at the current time (if $\lambda$ is too large) \,\cite{RatraPeebles88,Copeland98}. This can be cured with a sum of two exponentials with different values of $\lambda$\,\cite{Barreiro2000}, but of course it is less motivated since involves more parameters. Thus, the PR-potential seems to have the minimal number of ingredients to successfully accomplish the job. In point of fact, it is what we have now verified at a rather significant confidence level in the light of the modern cosmological data.

Finally, let us mention that we have tested the robustness of our computational program by setting the initial conditions out of the tracker path and recovering the asymptotic attractor trajectory. This is of course a numerical check, which is nicely consistent with the fact that the Peebles \& Ratra potential satisfies the aforementioned tracker condition $\Gamma>1$. More details will be reported elsewhere.

\begin{figure}
\begin{center}
\label{Parella2}
\includegraphics[width=5.0in, height=2.32in]{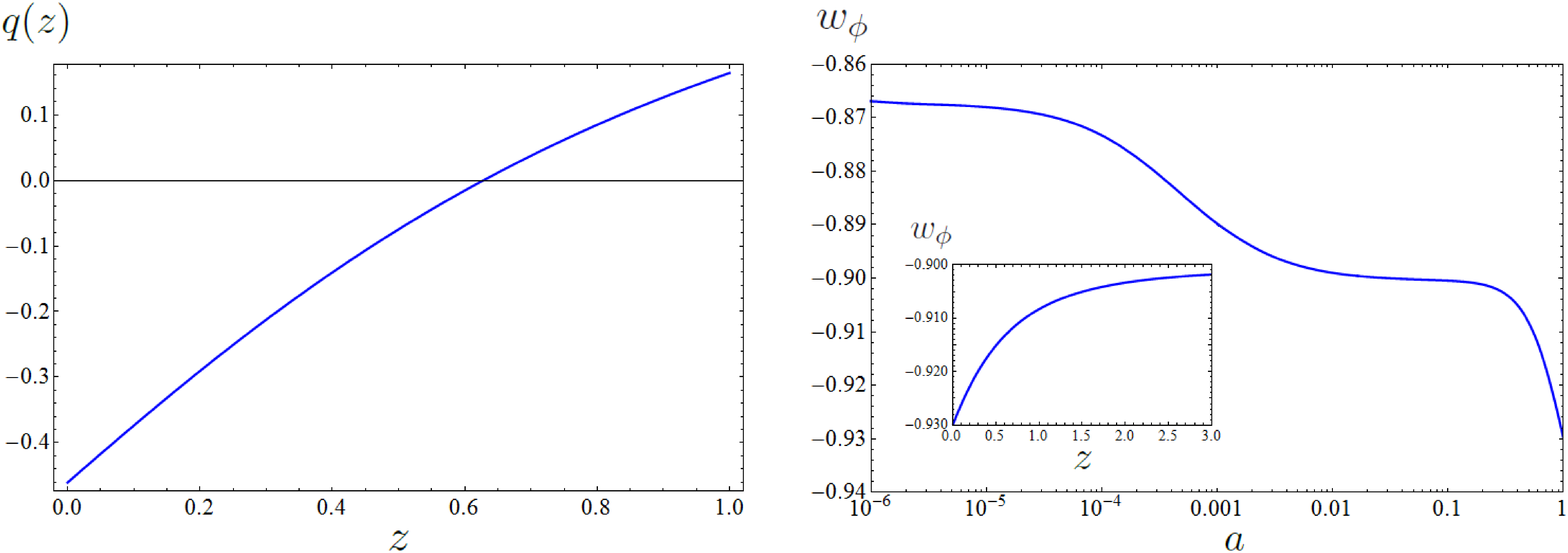}
\caption{\scriptsize
{\bf Left:} The deceleration parameter $q(z)$ for the recent Universe. The transition point where $q(z_t)=0$ is at $z_t=0.628$, for the best fit values of Table 2. {\bf Right:} The scalar field EoS parameter, $w_{\phi}(a)$, for the entire cosmic history after numerically solving the cosmological equations for the $\phi$CDM model with Peebles \& Ratra potential using the best-fit values of Table 2. The two plateaus from left to right correspond to the epochs of radiation and matter domination, respectively. The sloped stretch at the end, which is magnified in the inner plot in terms of  the redshift variable, corresponds to the recent epoch, in which the scalar field density (playing the role of DE) dominates. We find  $w_\phi(z=0)= -0.931\pm 0.017$.}
\end{center}
\end{figure}

%%%%%%%%%%%%%%%%%%%%%%%%%%%%%%%%%%%%%%%%%%%%%%%%%%%%%%%%%%%%%%%%%%%%%%%%%%%%%%%%%%%%%%%%%%%%%%%%%%%%%%%%
\section{XCDM and CPL parametrizations}
\label{sect:XCDMand CPL}
The XCDM parametrization was first introduced in\,\cite{XCDM} as the simplest way to track a possible dynamics for the DE. Here one replaces the $\CC$-term with an unspecified dynamical entity $X$, whose energy density at present coincides with the current value of the vacuum energy density, i.e. $\rho_X^0=\rLo$. Its EoS reads $p_X=w_0\,\rho_X$, with $w_0=$const. The XCDM mimics the behavior of a scalar field, whether quintessence ($w_0\gtrsim-1$) or phantom ($w_0\lesssim-1$), under the assumption that such field has an essentially constant EoS parameter around $-1$.  Since both matter and DE are self-conserved in the XCDM (i.e. they are not interacting), the energy densities as a function of the scale factor are given by $\rho_m(a)=\rho_m^0\,a^{-3}=\rho_{c 0}\Omega_m\,a^{-3}$ and $\rho_X(a)=\rho_X^0\,a^{-3(1+w_0)}=\rho_{c 0}(1-\Omega_m)\,a^{-3(1+w_0)}$.
Thus, the Hubble function in terms of the scale factor is given by
\begin{equation}\label{eq:HXCDM}
H^2(a)=%\frac{8\pi G}{3}\left[\rho_m^0\,a^{-3}+\rho_X^0\,a^{-3(1+\omega_0)}\right]=
H_0^2\left[\Omega_m\,a^{-3}+(1-\Omega_m)\,a^{-3(1+w_0)}\right]\,.
\end{equation}
A step further in the parametrization of the DE is the CPL prametrization\,\cite{CPL1,CPL2,CPL3}, whose EoS for the DE is defined as follows:
\begin{equation}\label{eq:CPL}
w=w_0+w_1\,(1-a)=w_0+w_1\,\frac{z}{1+z}\,,
\end{equation}
where $z$ is the cosmological redshift.
In contrast to the XCDM, the EoS of the CPL is not constant and is designed to have a well-defined asymptotic limit in the early Universe. The XCDM serves as a simple baseline to compare other models for the dynamical DE. The CPL further shapes the XCDM parametrization at the cost of an additional parameter ($w_1)$ that enables some cosmic evolution of the EoS. The Hubble function for the CPL in the MDE is readily found:
\begin{eqnarray}
\label{Hzzzquint} H^2(z)&=&
H_0^2\,\left[\Omega_m\,(1+z)^3+(1-\Omega_m)
(1+z)^{3(1+w_0+w_1)}\,e^{-3\,w_1\,\frac{z}{1+z}}\right]
 \,.
\end{eqnarray}

%%%%%%%%%%%%%%%%%%%%%%%%%%%%%%%%%%%%%%%%%%%%%%%%%%%%%%%%%%%%%%%%%%%%%%%%%%%%%%%%%%
\begin{figure}
\begin{center}
\label{Parella3}
\includegraphics[width=5.0in, height=2.32in]{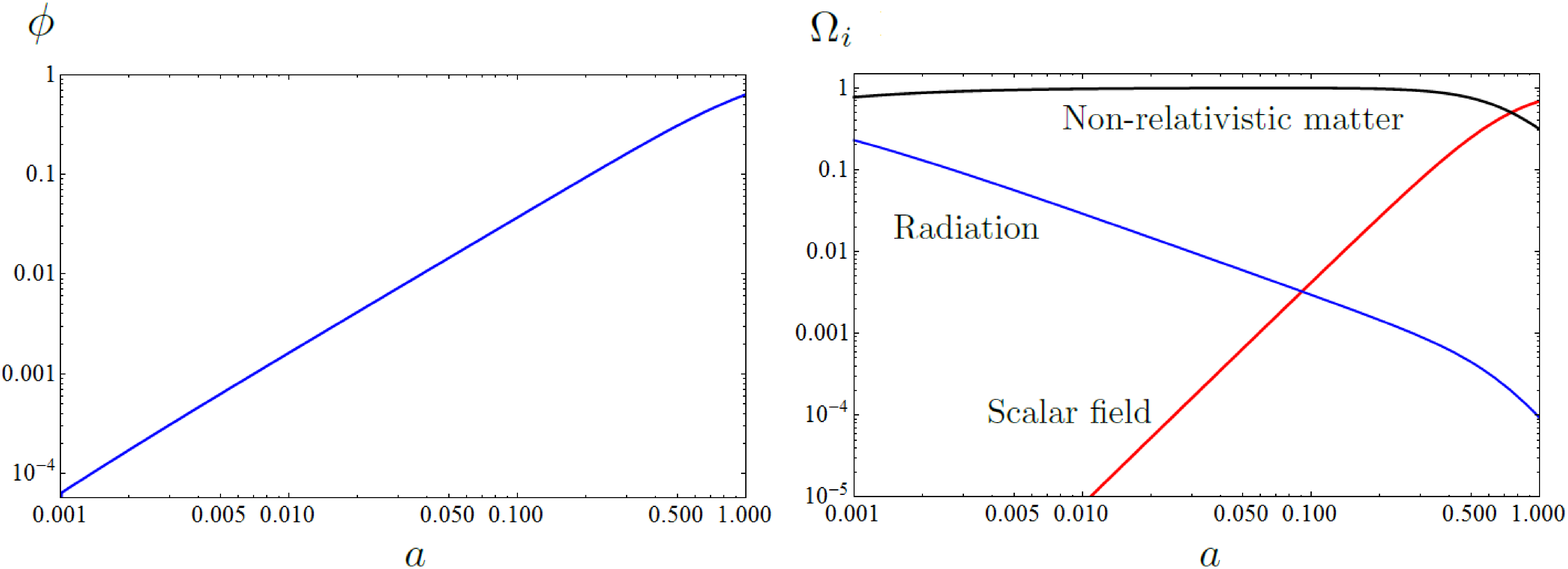}
\caption{\scriptsize Same as Fig.\,1, but for $\phi(a)$ and the density parameters $\Omega_i(a)$.  The crossing point between the scalar field density and the non-relativistic matter density lies very close to our time (viz. $a_c=0.751$, equivalently $z_c=0.332$), as it should. This is the point where the tail of the PR potential becomes visible and appears in the form of DE. The point $z_c$ lies nearer our time than the transition redshift from deceleration to acceleration, $z_t=0.628$ (cf. $q(z)$ in Fig. 1), similar to the $\CC$CDM.}
\end{center}
\end{figure}
%%%%%%%%%%%%%%%%%%%%%%%%%%%%%%%%%%%%%%%%%%%%%%%%%%%%%%%%%%%%%%%%%%%%%%%%%%%%%%%
It boils down to (\ref{eq:HXCDM}) for $w_1=0$, as expected. It is understood that for  the RDE  the term $\Omega_r (1+z)^4$ has to be added in the Hubble function. Such radiation term is already relevant for the analysis of the CMB data, and it is included in our analysis.
The fitting results for the XCDM and CPL parametrizations have been collected in the first two rows of Table 3.
Comparing with the $\phi$CDM model (cf. Table 2), we see that the XCDM parametrization also projects the effective quintessence option  at $4\sigma$ level, specifically $w_0=-0.916\pm0.021$.  The CPL parametrization, having one more parameter, does not reflect the same level of significance, but the corresponding AIC and BIC parameters (see below) remain relatively high as compared to the $\CC$CDM, therefore pointing also at clear signs of dynamical DE as the other models. Although the results obtained by the XCDM parametrization and the PR-model are fairly close (see Tables 2 and 3) and both EoS values  lie in the quintessence region, the fact that the EoS of the XCDM model is constant throughout the cosmic history makes it difficult to foresee if the XCDM can be used as a faithful representation of a given nontrivial $\phi$CDM model, such as the one we are considering here. The same happens for the extended CPL parametrization, even if in this case the EoS has some prescribed cosmic evolution. In actual fact, both the XCDM and CPL parametrizations are to a large extent arbitrary and incomplete representations of the dynamical DE.

\section{RVM: running vacuum}
\label{sect:RVM}
The last model whose fitting  results are reported in Table 3 is the running vacuum model (RVM). We  provide here the basic definition of it and some motivation -- see \,\cite{JSPRev2013,SolGom2015,JSPRev2016} and references therein for details.  The RVM is a dynamical vacuum model, meaning that the corresponding EoS parameter is still $w=-1$ but the corresponding vacuum energy density is a``running'' one, i.e. it departs  (mildly) from the rigid assumption $\rL=$const. of the $\CC$CDM. Specifically, the form of $\rL$ reads as follows:
\begin{equation}\label{eq:RVMvacuumdadensity}
\rho_\CC(H) = \frac{3}{8\pi{G}}\left(c_{0} + \nu{H^2}\right)\,.
\end{equation}
Here $c_0=H_0^2\left(1-\Omega_m-\nu\right)$ is fixed by the boundary condition $\rL(H_0)=\rLo=\rho_{c0}\,(1-\Omega_m)$. The dimensionless coefficient $\nu$ is expected to be very small, $|\nu|\ll1$, since the model must remain sufficiently close to the $\CC$CDM. The moderate dynamical evolution of $\rL(H)$ is possible at the expense of the slow decay rate of vacuum into dark matter (we assume that baryons and radiation are conserved\,\cite{DVM,JSPRev2016}).

In practice, the confrontation of the RVM with the data is performed by means of the following $5$-dimensional fitting vector:
\begin{equation}\label{eq:vfitting}
\vec{p}_{\rm RVM}=(\Omega_m,\omega_b,n_s, h,
\nu)\,.
\end{equation}
The first four parameters are the standard ones as in the $\CC$CDM, while $\nu$ is the mentioned vacuum parameter for the RVM. Although it can be treated in a mere phenomenological fashion,
formally $\nu$ can be given a QFT meaning by linking it to the $\beta$-function of the running $\rL$\,\cite{JSPRev2013,SolGom2015}. Theoretical estimates place its value in the ballpark of $\nu\sim 10^{-3}$ at most\,\cite{Fossil07}, and this is precisely the order of magnitude pinned down for it in Table 3 from our overall fit to the data. The order of magnitude coincidence is reassuring.
Different realizations of the RVM are possible \cite{ApJL,RVM,DVM,JSPRev2016}, but here we limit ourselves to the simplest version. The corresponding Hubble function in the MDE reads:
\begin{equation}\label{eq:H2RVM}
H^2(z)=H_0^2\,\left[1+\frac{\Omega_m}{1-\nu}\left((1+z)^{3(1-\nu)}-1\right)\right]\,.
\end{equation}
It depends on the basic fitting parameters $(\Omega_m, h,\nu)$, which are the counterpart of $(\omega_m,\alpha,\bar{\kappa})$ for the $\phi$CDM. The remaining two parameters are common and hence both for the RVM and the $\phi$CDM the total number of fitting parameter is five, see (\ref{eq:vfittingPhiCDM}) and (\ref{eq:vfitting}).
Note that for $\nu=0$ we recover the $\CC$CDM case, as it should be expected.

%%%%%%%%%%%%%%%%%%%%%%%%%%%%%%%%%%%%%%%%%%%%%%%%%%%%%%%%%%%%%%%%%%%%%%%%%%%%%%%

\begin{figure}
\begin{center}
\label{Parella1}
\includegraphics[width=3.5in, height=2.4in]{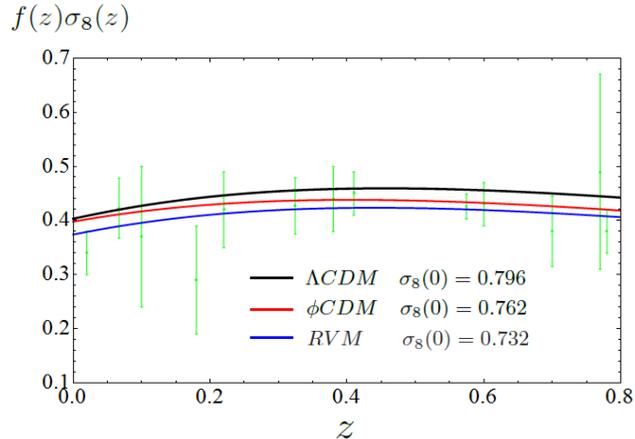}
\caption{\scriptsize The LSS data on the weighted linear growth rate, $f(z)\sigma_8(z)$, and the predicted curves by the various models, using the best-fit values in Tables 1-3. The XCDM and CPL lines are not shown since they are almost on top of the $\phi$CDM one. The values of $\sigma_8(0)$ that we obtain for the different models are also indicated.
}
\end{center}
\end{figure}

%%%%%%%%%%%%%%%%%%%%%%%%%%%%%%%%%%%%%%%%%%%%%%%%%%%%%%%%%%%%%%%%%%%%%%%%%%%%%%%%%%%%

\section{Structure formation}
A few observations on the analysis of structure formation are in order, as it plays a significantly role in the fitting results. On scales well below the horizon the scalar field perturbations are relativistic and hence can be ignored\,\cite{PeeblesRatra88}. As a result in the presence of non-interacting scalar fields the usual matter perturbation equation remains valid\,\cite{DEBook}. Thus, for the $\phi$CDM, XCDM and CPL models we compute the perturbations through the standard equation\,\cite{Peebles1993}
\begin{equation}\label{diffeqLCDM}
\ddot{\delta}_m+2H\,\dot{\delta}_m-4\pi
G\rmr\,\delta_m=0\,,
\end{equation}
with, however, the Hubble function corresponding to each one of these models -- see the formulae in the previous sections.

For the RVM the situation is nevertheless different. In the presence of dynamical vacuum, the perturbation equation not only involves the modified Hubble function (\ref{eq:H2RVM}) but the equation itself becomes modified. The generalized perturbation equation reads\,\cite{GomSolBas2015,AGVsola2015,AGVsolKa2015}:
\begin{equation}\label{diffeqD}
\ddot{\delta}_m+\left(2H+\Psi\right)\,\dot{\delta}_m-\left(4\pi
G\rmr-2H\Psi-\dot{\Psi}\right)\,\delta_m=0\,,
\end{equation}
where $\Psi\equiv -\dot{\rho}_{\Lambda}/{\rmr}$. As expected, for $\rL=$const. we have $\Psi=0$ and Eq.\,(\ref{diffeqD}) reduces to the standard one (\ref{diffeqLCDM}).
To solve the above perturbation equations we have to fix the initial conditions for $\delta_m$ and $\dot{\delta}_m$ for each model at high redshift, say at $z_i\sim100$ ($a_i\sim10^{-2}$), when non-relativistic matter dominates over the vacuum -- confer Ref.\,\cite{GomSolBas2015,AGVsola2015,AGVsolKa2015}.

Let us also note that, in all cases, we can neglect the DE perturbations at subhorizon scales. We have already mentioned above that this is justified for the $\phi$CDM. For the RVM it can be shown to be also the case, see \,\cite{GomSolBas2015,AGVsola2015,AGVsolKa2015}. The situation with the XCDM and CPL is not different, and once more the DE perturbations are negligible at scales below the horizon. A detailed study of this issue can be found e.g. in Ref.\,\cite{LXCDM1,LXCDM2}, in which the so-called $\CC$XCDM model is considered in detail at the perturbations level. In the absence of the (running) component $\CC$ of the DE, the $\CC$XCDM model reduces exactly to the XCDM as a particular case. One can see in that quantitative study that at subhorizon scales the DE perturbations become negligible no matter what is the  assumed value for the sound velocity of the DE perturbations (whether adiabatic or non-adiabatic).

The analysis of the linear LSS regime is conveniently implemented with the help of the weighted linear growth $f(z)\sigma_8(z)$, where $f(z)=d\ln{\delta_m}/d\ln{a}$ is the growth factor and $\sigma_8(z)$ is the rms mass fluctuation amplitude on scales of $R_8=8\,h^{-1}$ Mpc at redshift $z$. It is computed as follows:
\begin{equation}
\begin{small}\sigma_{\rm 8}(z)=\sigma_{8, \CC}
\frac{\delta_m(z)}{\delta^{\CC}_{m}(0)}
\sqrt{\frac{\int_{0}^{\infty} k^{n_s+2} T^{2}(\vec{p},k)
W^2(kR_{8}) dk} {\int_{0}^{\infty} k^{n_{s,\CC}+2} T^{2}(\vec{p}_\Lambda,k) W^2(kR_{8,\Lambda}) dk}}\,,\label{s88general}
\end{small}\end{equation}
where $W$ is a top-hat smoothing function and $T(\vec{p},k)$ the transfer function (see e.g. \cite{GomSolBas2015,AGVsola2015,AGVsolKa2015} for details). Here $\vec{p}$ stands for the corresponding fitting vector for the various models, as indicated in the previous sections. In addition, we define a fiducial model, which we use in order to fix the normalization of the power spectrum. For that model we take the $\CC$CDM at fixed parameter values from the Planck 2015 TT,TE,EE+lowP+lensing analysis\,\cite{Planck2015}. Such fiducial values are collected in the vector
$\vec{p}_\CC=(\Omega_{m,\CC},\omega_{b,\CC},n_{s,\CC},h_{\CC})$.
In Fig. 3 we display  $f(z)\sigma_8(z)$ for the various models using the fitted values of Tables 1-3 following this procedure.

%%%%%%%%%%%%%%%%%%%%%%%%%%%%%%%%%%%%%%%%%%%%%%%%%%%%%%%%%%%%%%%%%%%%%%%%%%%%%%%
\begin{figure}
\begin{center}
\label{Parella2b}
\includegraphics[width=4.0in, height=2.2in]{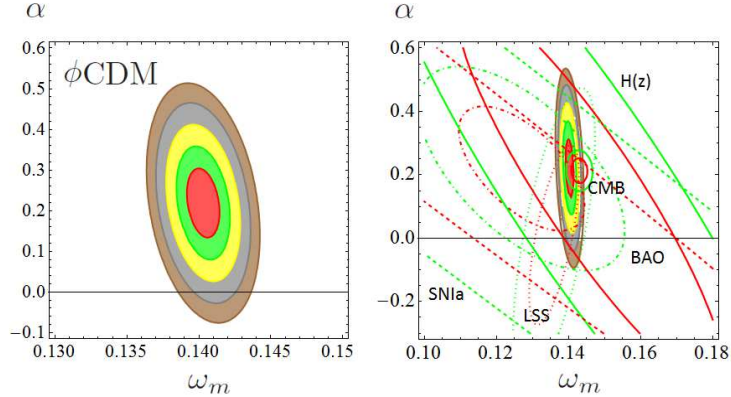}
\caption{\scriptsize {\bf Left}: Likelihood contours for the $\phi$CDM model in the ($\omega_m$,$\alpha$)-plane after marginalizing over the remaining parameters (cf. Table 2). The various contours correspond to 1$\sigma$, 2$\sigma$, 3$\sigma$, 4$\sigma$ and 5$\sigma$  c.l. The line $\alpha=0$ corresponds to the concordance $\CC$CDM model. The tracker consistency region $\alpha>0$ (see the text) is clearly preferred, and we see that it definitely points to dynamical DE at $\sim4\sigma$ confidence level. {\bf Right}: Reconstruction of the aforementioned contour lines from the partial contour plots of the different SNIa+BAO+$H(z)$+LSS+CMB data sources using Fisher's approach\,\cite{AmendolaStatistics}. The $1\sigma$ and $2\sigma$ contours are shown in all cases, but for the reconstructed final contour lines we include the $3\sigma$, $4\sigma$ and $5\sigma$ regions as well. For the reconstruction plot we display a larger $\omega_m$-range to better appraise the impact of the various data sources.}
\end{center}
\end{figure}

\begin{figure}
\begin{center}
\label{Parella3b}
\includegraphics[width=3.5in, height=2.11in]{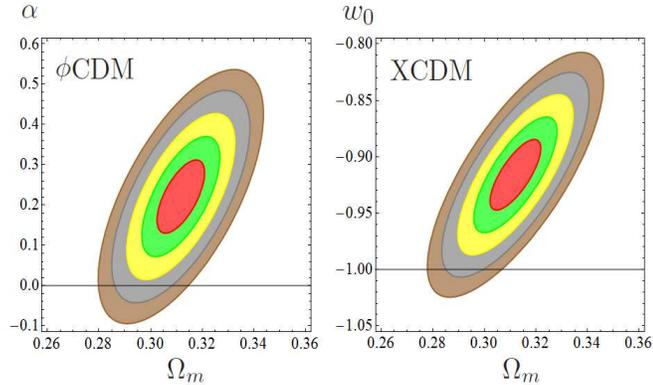}
\caption{\scriptsize Likelihood contours for the $\phi$CDM model (left) and the XCDM parametrization (right) in the relevant planes after marginalizing over the remaining parameters in each case (cf. Tables 2 and 3). The various contours correspond to 1$\sigma$, 2$\sigma$, 3$\sigma$, 4$\sigma$ and 5$\sigma$ c.l. The central values in both cases are $\sim4\sigma$ away from the $\CC$CDM, i.e. $\alpha=0$ and $w_0=-1$, respectively.}
\end{center}
\end{figure}

%%%%%%%%%%%%%%%%%%%%%%%%%%%%%%%%%%%%%%%%%%%%%%%%%%%%%%%%%%%%%%%%%%%%%%%%%%%%%%%%

\section{Discussion and conclusions}
The statistical analysis of the various models considered in this study is performed in terms of a joint likelihood function, which is the product of the likelihoods for each data source
%\begin{equation}\label{eq:chi2}
%\chi^2_{tot}=\chi^2_{SNIa}+\chi^2_{BAO}+\chi^2_{H}+\chi^2_{f\sigma_8}+\chi^2_{CMB}\,.
%\end{equation}
and including the corresponding covariance matrices, following the standard procedure\,\cite{DEBook,AmendolaStatistics}. The contour plots for the $\phi$CDM and XCDM models are shown in Figures 4 and 5, where the the dynamical character of the DE is clearly demonstrated at $\sim 4\sigma$ c.l. More specifically, in the left plot of  Fig. 4 we display the final contour plots for $\phi$CDM in the plane $(\omega_m,\alpha)$ -- defined by two of the original parameters of our calculation, cf. Eq.\,(\ref{eq:vfittingPhiCDM}) --  together with the isolated contours of the different data sources (plot on the right). It can be seen that the joint triad of observables BAO+LSS+CMB conspire to significantly reduce the final allowed region of the ($\omega_m,\alpha$)-plane, while the constraints imposed by SNIa and $H(z)$ are much weaker. Finally, for the sake of convenience, in Fig. 5 we put forward the final $\phi$CDM and the XCDM contours in the more conventional $(\Omega_m,\alpha)$-plane. As for the RVM, see the contours in \cite{RVM,DVM} and \cite{JSPRev2016}, where a dynamical DE effect $\gtrsim 4\sigma$  is recorded.

As noted previously, the three models $\phi$CDM, XCDM and RVM have the same number of parameters, namely 5, one more than the $\CC$CDM. The CPL, however, has 6 parameters. Cosmological models having a larger number of parameters have more freedom to accommodate observations. Thus, for a fairer comparison of the various nonstandard models with the concordance $\CC$CDM we have to invoke a suitable statistical procedure that penalizes the presence of extra parameters. Efficient criteria of this kind are available in the literature and they have been systematically used in different contexts to help making a selection of the best candidates among competing models describing the same data.  For a long time it has been known that the Akaike information criterion (AIC) and the Bayesian information criterion (BIC) are extremely valuable tools for a fair statistical analysis of this kind. These criteria are defined as follows\,\cite{Akaike,Schwarz,Burnham}:
\begin{equation}\label{eq:AICandBIC}
{\rm AIC}=\chi^2_{\rm min}+\frac{2nN}{N-n-1}\,,\ \ \ \ \
{\rm BIC}=\chi^2_{\rm min}+n\,\ln N\,,
\end{equation}
where $n$ is the number of independent fitting parameters and $N$ the number of data points.
The larger are the differences $\Delta$AIC ($\Delta$BIC) with respect to the model that carries smaller value of AIC (BIC) the higher is the evidence against the model with larger value of  AIC (BIC) -- the $\CC$CDM in all the cases considered in Tables 2-3.
The rule applied to our case is the following\,\cite{Akaike,Schwarz,Burnham}: for $\Delta$AIC and $\Delta$BIC in the range $6-10$ we can speak of ``strong evidence'' against the $\CC$CDM, and hence in favor of the given nonstandard model. Above 10, one speaks of ``very strong evidence''. Notice that the Bayes factor is $e^{\Delta {\rm BIC}/2}$, and hence near 150 in such case.

A glance at Tables 2 and 3 tells us that for the models $\phi$CDM, XCDM and RVM, the values of $\Delta$AIC and $\Delta$BIC are both above 10. The CPL parametrization has only one of the two increments above 10, but the lowest one is above 6, therefore it is still fairly (but not so strongly) favored as the others. We conclude from the AIC and BIC criteria that the models $\phi$CDM, XCDM and RVM  are definitely selected over the $\CC$CDM as to their ability to optimally fit the large set of cosmological SNIa+BAO+$H(z)$+LSS+CMB data used in our analysis.  Although the most conspicuous model of those analyzed here appears to be the RVM {(cf. Tables 2 and 3)}, the scalar field model $\phi$CDM with Peebles \& Ratra potential also receives a strong favorable verdict from the AIC and BIC criteria. Furthermore, the fact that the generic XCDM and CPL parametrizations are also capable of detecting significant signs of evolving DE suggests that such dynamical signature is sitting in the data and is not privative of a particular model, although the level of sensitivity does indeed vary from one model to the other.

To summarize, the current cosmological data disfavors the hypothesis $\CC=$const. in a rather significant way. The presence of DE dynamics is confirmed by all four parametrizations considered here and with a strength that ranges between strong and very strong evidence, according to the Akaike and Bayesian information criteria. Furthermore, three of these parametrizations are able to attest such evidence at $\sim4\sigma$ c.l., and two of them ($\phi$CDM and RVM) are actually more than parametrizations since they are associated to specific theoretical frameworks.  The four approaches resonate in harmony with the conclusion that the DE is decreasing with the expansion, and therefore that it behaves effectively as quintessence.

%%%%%%%%%%%%%%%%%%%%%%%%%%%%%%%%%%%%%%%%%%%%%%%%%%%%%%%%%%%%%%%%%
%%%%%%%%%%%%%%%%%%%%%%%%%%%%%%%%%%%%%%%%%%%%%%%%%%%%%%%%%%%%%%%%%
%%%%%%%%%%%%%%%%%%%%%%%%%%%%%%%%%%%%%%%%%%%%%%%%%%%%%%%%%%%%%%%%%

\section{Acknowledgements}
\noindent JS has been supported by FPA2013-46570 (MICINN), Consolider
grant CSD2007-00042 (CPAN) and by 2014-SGR-104 (Generalitat de
Catalunya); AGV acknowledges the support of an APIF  grant of the
U. Barcelona. We are partially supported by  MDM-2014-0369 (ICCUB).

%%%%%%%%%%%%%%%%%%%%%%%%%%%%%%%%%%%%%%%%%%%%%%%%%%%%%%%%%%%%%%%%%
%%%%%%%%%%%%%%%%%%%%%%%%%%%%%%%%%%%%%%%%%%%%%%%%%%%%%%%%%%%%%%%%%
%%%%%%%%%%%%%%%%%%%%%%%%%%%%%%%%%%%%%%%%%%%%%%%%%%%%%%%%%%%%%%%%%

\newcommand{\CQG}[3]{{ Class. Quant. Grav. } {\bf #1} (#2) {#3}}
\newcommand{\JCAP}[3]{{ JCAP} {\bf#1} (#2)  {#3}}
\newcommand{\APJ}[3]{{ Astrophys. J. } {\bf #1} (#2)  {#3}}
\newcommand{\AMJ}[3]{{ Astronom. J. } {\bf #1} (#2)  {#3}}
\newcommand{\APP}[3]{{ Astropart. Phys. } {\bf #1} (#2)  {#3}}
\newcommand{\AAP}[3]{{ Astron. Astrophys. } {\bf #1} (#2)  {#3}}
\newcommand{\MNRAS}[3]{{ Mon. Not. Roy. Astron. Soc.} {\bf #1} (#2)  {#3}}
\newcommand{\PR}[3]{{ Phys. Rep. } {\bf #1} (#2)  {#3}}
\newcommand{\RMP}[3]{{ Rev. Mod. Phys. } {\bf #1} (#2)  {#3}}
\newcommand{\JPA}[3]{{ J. Phys. A: Math. Theor.} {\bf #1} (#2)  {#3}}
\newcommand{\ProgS}[3]{{ Prog. Theor. Phys. Supp.} {\bf #1} (#2)  {#3}}
\newcommand{\APJS}[3]{{ Astrophys. J. Supl.} {\bf #1} (#2)  {#3}}
%%%%%%%%%%%%%%%%%%%%%%%%%%%%%%%%%%%%%%%%%%%%%%%%%%%%%%%%%%%%%%%%%%%%%%%%%

\newcommand{\Prog}[3]{{ Prog. Theor. Phys.} {\bf #1}  (#2) {#3}}
\newcommand{\IJMPA}[3]{{ Int. J. of Mod. Phys. A} {\bf #1}  {(#2)} {#3}}
\newcommand{\IJMPD}[3]{{ Int. J. of Mod. Phys. D} {\bf #1}  {(#2)} {#3}}
\newcommand{\GRG}[3]{{ Gen. Rel. Grav.} {\bf #1}  {(#2)} {#3}}

%%%%%%%%%%%%%%%%%%%%%%%%%%%%%%%%%%%%%%%%%%%%%%%%%%%%%%%%%%%%%%%%%
%%%%%%%%%%%%%%%%%%%%%%%%%%%%%%%%%%%%%%%%%%%%%%%%%%%%%%%%%%%%%%%%%

\end{document}